\documentclass[10pt,twocolumn]{article}
\usepackage[margin=0.7in]{geometry}
\usepackage{lineno}
\usepackage{authblk}
\usepackage{amsmath}
\usepackage{float}
\usepackage{graphicx}
\usepackage[utf8]{inputenc}
\usepackage[english]{babel}
\usepackage[comma,numbers,sort&compress,super]{natbib} 

\usepackage[colorlinks=true, allcolors=blue]{hyperref}
\usepackage[font={footnotesize,it}]{caption}
\usepackage{kpfonts}

\title{\bfseries  Remote determination of the shape of Jupiter's vortices from laboratory experiments}

\author[a]{\bfseries  Daphné Lemasquerier\footnote{Corresponding author: lemasquerier.pro@protonmail.com}}
\author[a]{\bfseries  Giulio Facchini\footnote{Current affiliation: Life Sciences Department, University of Roehampton, Whitehands College, Holybourne ave., London, SW15 4JD, UK}} 
\author[a]{\bfseries  Benjamin Favier}
\author[a]{\bfseries  Michael Le Bars}

\affil[a]{Aix-Marseille Univ, CNRS, Centrale Marseille, Institut de Recherche sur les Ph{\'e}nomènes Hors {\'E}quilibre, UMR 7342, 49 rue F. Joliot Curie, 13013 Marseille, France}

\date{}

\begin{document}

\maketitle

\renewcommand{\abstractname}{\vspace{-\baselineskip}}

\begin{abstract}
\bfseries Jupiter’s dynamics shapes its cloud patterns but remains largely unknown below this natural observational barrier. Unraveling the underlying three-dimensional flows is thus a primary goal for NASA's ongoing Juno mission that was launched in 2011. Here, we address the dynamics of large Jovian vortices using laboratory experiments complemented by theoretical and numerical analyses. We determine the generic force balance responsible for their three-dimensional pancake-like shape. From this, we define scaling laws for their horizontal and vertical aspect ratios as a function of the ambient rotation, stratification and zonal wind velocity. For the Great Red Spot in particular, our predicted horizontal dimensions agree well with measurements at the cloud level since the Voyager mission in 1979. We additionally predict the Great Red Spot’s thickness, inaccessible to direct observation: it has surprisingly remained constant despite the observed horizontal shrinking. Our results now await comparison with upcoming Juno observations.
\end{abstract}

Earth-based telescope observations and records from spacecrafts -- including the ongoing \textit{Juno} mission \citep[][]{li_distribution_2017,kaspi_jupiters_2018,iess_measurement_2018,guillot_suppression_2018,adriani_clusters_2018,kong_origin_2018,debras_new_2019} -- have revealed Jupiter’s rich tropospheric dynamics. Among other salient features, several hundred vortices are embedded within Jupiter’s zonal winds \citep{vasavada_jovian_2005}, the most famous one being the Great Red Spot (GRS) observed for at least 100 years\cite{rogers_giant_1995} if not 350 years \citep{falorni_discovery_1987}. Yet, these vortices raise questions still discussed today: how do they form? What controls their lifetime? How do they interact with Jupiter's zonal flows? What is their three-dimensional structure, and more specifically their thickness? Are they columns that penetrate through the molecular envelope \citep{hide_origin_1961}, or shallow vortices confined near the cloud level \citep{marcus_numerical_1988,dowling_jupiters_1988}? Idealized numerical models \citep[][]{williams_stability_1988,dowling_potential_1988,marcus_numerical_1988,marcus_vortex_1990} and laboratory experiments \citep{read_long-lived_1983,read_isolated_1984,sommeria_laboratory_1988,antipov_rossby_1986} have offered clues to understand vortex formation, interaction and longevity in Jupiter's atmosphere, in complement to measurements. Here, we address the question of their still unknown depth, inaccessible to direct observation.

From a dynamical point of view, vortices naturally arise in planetary flows subjected to rapid rotation, owing to the so-called geostrophic equilibrium: the Coriolis term in the momentum equation balances the horizontal pressure gradient, and the flow rotates in opposite directions around low and high pressure zones. In the sole presence of rotation, the vortices are expected to extend vertically in columns  throughout the fluid layer owing to the Taylor-Proudman theorem. But in planetary flows, stratification often comes into play besides rotation: rather than columns, vortices take the shape of thin pancakes\citep{billant_self-similarity_2001}. For instance, mesoscale vortices in the Earth's ocean (the so-called meddies\citep{carton_hydrodynamical_2001}) are embedded in a strong and stable thermohaline stratification: their structure, as revealed through direct velocity, temperature and salinity measurements, exhibits a lenticular shape, with radii from 20 to 100 km and thicknesses lower than $\sim$ 1 km. Their aspect ratio depends on the meddy vorticity, the stratification difference between the vortex and the ocean, and the background rotation \citep{aubert_universal_2012,hassanzadeh_universal_2012}. 

Similarly, Jupiter's vortices lie in a stratified layer: the weather layer above the convective zone and below the tropopause \citep{vasavada_jovian_2005}. But contrary to oceanic vortices, direct measurements to investigate their three-dimensional shape are barely accessible, and observations remain limited to the cloud level. Additionally, Jovian vortices are embedded in strong zonal winds related to Jupiter's jets\citep{mitchell_flow_1981}. 

Inviscid and purely two-dimensional elliptical vortices embedded in a uniform strain have been studied extensively in non-rotating frameworks \citep{moore_structure_1971,kida_motion_1981,meacham_vortices_1989,moffatt_stretched_1994}. In rotating and/or stratified flows, but with no strain, studies were mainly dedicated to vortex stability \citep{meacham_quasigeostrophic_1992,sipp_vortices_1999,godeferd_zonal_2001,riedinger_instability_2010,yim_stability_2016}. Here, our goal is to investigate the three-dimensional equilibrium shape of a vortex in a medium where the three main planetary ingredients -- rotation, stratification and shear -- coexist.
To do so, we use an experimental setup which allows us to generate vortices in a model flow with the three aforementioned ingredients and to follow their temporal evolution. Experimental results are rationalized and extended by combined numerical and theoretical analyses of the generic equations of motion. Accounting for the facts that the vast majority ($\sim$90\%) of Jupiter's vortices are anticyclonic \citep{vasavada_jovian_2005} and that all long-lived Jovian vortices have relative vorticity with same sign as that of the shear in which they are embedded \citep{vasavada_jovian_2005}, we focus on anticyclones embedded in an anticyclonic shear.

\begin{figure}[t]
	\centering
	\includegraphics[width=1\linewidth]{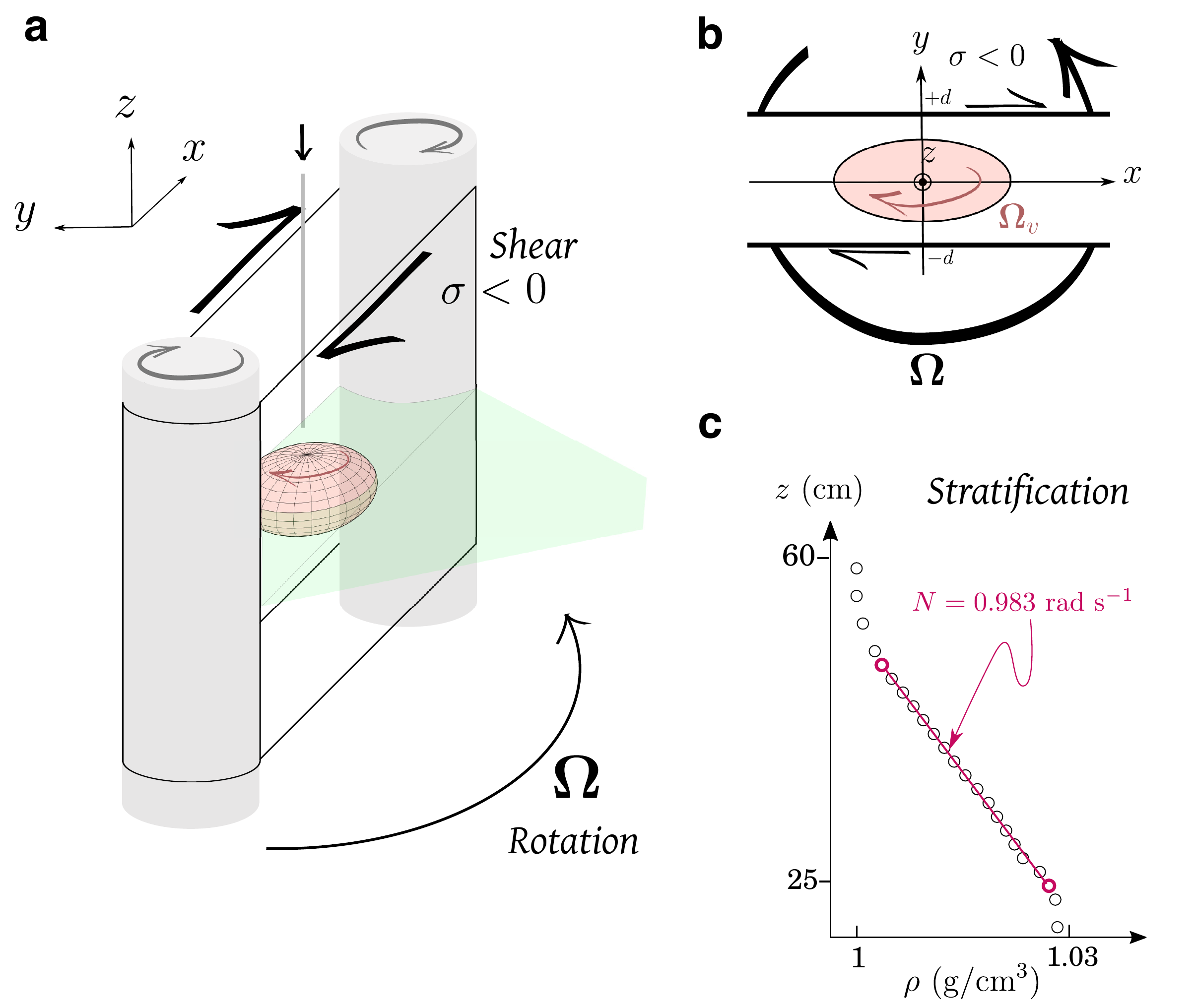}
	\caption{Simplified sketch of the experimental set-up. \textbf{a.}  The shearing device is made of a PVC belt encircling two co-rotating cylinders. A capillary tube injects fluid in the gap between the membrane sides to create an anticyclone which is analyzed by performing Particle Image Velocimetry (PIV) in its equatorial plane. The represented device is placed inside a bigger tank which rotates at a rate $\Omega$ and the fluid is stratified using salt water. \textbf{b.} Sketch of an equatorial view. A linear shear is added via the action of two rigid PVC boundaries separated by a distance $2d$. \textbf{c.} Example of a measured stratification, where $\rho$ is the dimensional density. The error bars are smaller than the markers. The red line is a linear fit used to determine the buoyancy frequency $N$ (equation \eqref{eq:buoyancy}). For all the experiments in the main text, $N = 1\pm0.1~{\rm rad~s^{-1}}$ .}
	\label{fig:vtx-setup}
\end{figure}

As sketched in Fig.\ref{fig:vtx-setup}, we consider the flow of an incompressible fluid of constant kinematic viscosity $\nu$ rotating around the vertical $z$-axis (oriented upward) at a constant rate $\boldsymbol{\Omega}=\Omega~\boldsymbol{e}_z$ (Coriolis frequency $f=2\Omega$). This flow is stably stratified, and characterized by its buoyancy frequency $N$ which is the natural frequency of oscillation of a fluid parcel displaced from equilibrium with buoyancy acting as a restoring force:
\begin{linenomath*}
	\begin{equation}
	N = \sqrt{- \frac{g}{\rho} \frac{\partial \rho}{\partial z}},
	\label{eq:buoyancy}
	\end{equation}
\end{linenomath*}
where $g$ is the gravitational acceleration and $\rho$ the density. The generic equations describing the flow are the continuity and Navier-Stokes equations in the Boussinesq approximation, as well as the advection-diffusion equation of the stratifying agent of constant diffusivity $\kappa$, whose concentration is linearly related to the density (see details in Supplementary Information section 1). In our experiments, the working fluid is salt water. A linear shear is added via the action of two rigid boundaries located at $y=(-d,d)$ moving at constant velocity in opposite directions parallel to $x$ (Fig.\ref{fig:vtx-setup}). The shear rate normalized by $f$ is denoted $\sigma$. 

In the dissipationless and linear limit and assuming a steady cyclo-geostrophic and hydrostatic equilibrium state (see details in Supplementary Information section 1), the equations of motion admit a solution under the form of a compact ellipsoidal vortex of constant vertical vorticity $\omega_c$.{ We define the corresponding Rossby number of the vortex as $Ro=\omega_c/2f$ ($Ro<0$ for anticyclones, and $>0$ for cyclones). Denoting $a$ and $b$ the vortex semi-major and semi-minor axes in an horizontal plane, and $c$ its vertical semi-axis, the corresponding velocity field in cartesian coordinates can be written as
\begin{linenomath*}
	\begin{equation}
	\boldsymbol{u_v} = Ro 
	\left(  
	\begin{array}{c}
	-(1+\beta)y\\
	(1-\beta)x\\
	0\\
	\end{array}
	\right),
	\label{eq:vtx-elliptic}	
	\end{equation}
\end{linenomath*}
where $\beta = (a^2-b^2)/(a^2+b^2)$ is the equatorial ellipticity of the vortex which goes from 0 for an axisymmetric vortex to 1 for an infinitely stretched ellipse. The stratification inside the vortex is assumed to be linear with a buoyancy frequency $N_c$. Continuity of the pressure field between the vortex and the surrounding imposed plane Couette flow of shear rate $\sigma$ defines the ellipsoidal contour of the vortex as 
\begin{linenomath*}
	\begin{eqnarray}
	& & Ro ~(1-\beta)~ [ 1+(1+\beta)Ro ]~ x^2 \nonumber \\
	+ && ~ \big( Ro ~(1+\beta)~ [ 1+(1-\beta)Ro ] - \sigma \big)~ y^2 \nonumber \\
	+ && ~ \frac{\big( N_c^2 - N^2\big)}{f^2} ~ z^2 = {\rm constant}.
	\label{eq:vtx-ellipsoid}
	\end{eqnarray}
\end{linenomath*}
Applying this relation at the points $(x,y,z)=(a,0,0)$ and $(0,b,0)$ and equating the two values give the relation
\begin{linenomath*}
	\begin{equation}
	\beta^2 \left( 2 \frac{Ro_x^2}{\sigma} +1 \right) + 2\beta  \left( \frac{Ro_x^2}{\sigma} -1 \right) + 1 = 0,
	\label{eq:vtx-horizaspect}
	\end{equation}
\end{linenomath*}   
where $Ro_x=(1-\beta)Ro$ is the streamwise Rossby number, that is the slope of the cross-stream velocity profile along $x$ at the vortex center. Knowing the strength of the vortex and the shear applied to it, this relation predicts the equatorial ellipticity of the vortex. From this equation, we select the root $\beta$ that is positive and comprised between 0 and 1. The ellipticity then evolves intuitively: for a weak ambient shear compare to the vortex intensity (i.e. $Ro_x^2/\vert\sigma\vert \gg 1$), the vortex tends towards axisymmetry ($\beta \rightarrow 0$). On the contrary, when $Ro_x^2/\vert\sigma\vert \ll 1$, $\beta=1$, meaning that the vortex is infinitely extended in the stream-wise direction ($a/b\gg1$).

Applying equation (\ref{eq:vtx-ellipsoid}) at $(x,y,z)=(a,0,0)$ and $(0,0,c)$ gives the vertical aspect ratio of the vortex:
\begin{linenomath*}
	\begin{equation}
	\left(\frac{c}{a}\right)^2 = \frac{Ro_x  \left[  1 + Ro_x~\frac{1+\beta}{1-\beta} \right] f^2 }{N_c^2 - N^2}. 
	\label{eq:vtx-vertaspect}
	\end{equation}
\end{linenomath*}
Interestingly, the shear does not directly appear in this relation, even if its influence is hidden in the ellipticity $\beta$. Thus, knowing only the horizontal aspect ratio, the strength of the vortex and its stratification, one can infer its vertical aspect ratio. For an axisymmetric vortex, i.e. without shear, $\beta=0$ and we retrieve the relation in the sole presence of rotation and stratification \cite{aubert_universal_2012,hassanzadeh_universal_2012}. On the contrary when $\beta \rightarrow 1$, since $Ro_x=(1-\beta)Ro$, the vortex is infinitely sheared and flat ($c/a \rightarrow 0$).

\begin{figure}[t]
	\centering
	\includegraphics[width=1\linewidth]{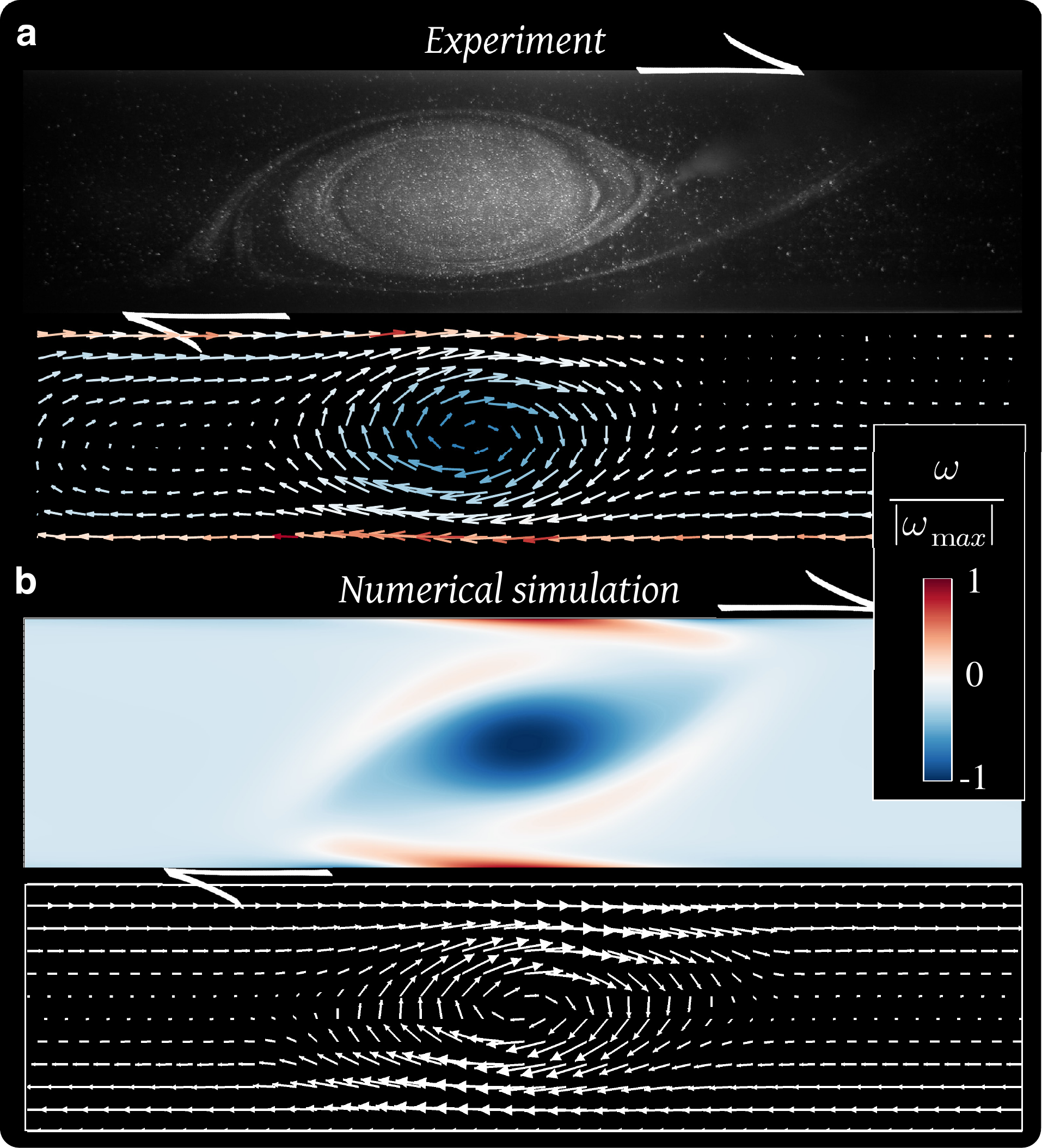}
	\caption{Top-view visualizations and corresponding velocity field in the vortex equatorial plane during an experiment and a numerical simulation. \textbf{a.} Experimental visualization of the vortex dyed with Rhodamine B. The associated velocity field is computed by PIV (one grid point out of three is kept in both directions for clarity). \textbf{b.} Snapshot from a numerical simulation representing the vertical component of the vorticity $\omega$ and the associated velocity field interpolated on a cartesian grid.}
	\label{fig:PIVfields}
\end{figure}

We now validate our theoretical model with both laboratory experiments and direct numerical simulations, where we follow the dynamical evolution of a single vortex through time $t$. Fig. \ref{fig:PIVfields} shows typical visualizations in the equatorial plane, with the corresponding velocity fields. Details about the experimental set-up and numerical approach are given in the Methods section and in the Supplementary Information, sections 2-4. We also show in section 5 that the dominant physical balances during the vortex evolution are consistent with the hypotheses assumed to derive the equilibrium shape: the vortex is at any time ellipsoidal, and the cyclo-geostrophic and hydrostatic equilibria are dominant. We now focus on the shape evolution of our laboratory and numerical vortices and compare the measurements with the theoretical laws (\ref{eq:vtx-horizaspect}) and (\ref{eq:vtx-vertaspect}). 
The evolution of the measured horizontal aspect ratio $a/b$ is represented as a function of $Ro_x/\vert \sigma \vert^{1/2}$ in Fig.\ref{fig:expnum-shape-evol}\textit{a} for five simulations and five experiments with different shear rates. At any time during the simulations and experiments, there is a good agreement between the measured equatorial shape of the vortex and our prediction. Fig.\ref{fig:expnum-shape-evol}\textit{b} shows the measured vertical aspect ratio $c/a$ as a function of the theoretical one. To compute the theoretical vertical aspect ratio, $Ro_x(t)$ and $\beta(t)$ are measured at each time. It is also necessary to know the internal stratification of the vortex $N_c(t)$. We have access to it numerically, but not experimentally. Thus, we use the approximation that the stratification does not change inside the vortex, that is $\forall t,~N_c(t)=0$ (fully mixed interior), even if the numerical results show that the density anomaly diminishes with time due to its advection by secondary, internal recirculation (see Supplementary Information, section 5.2 and Figure S5). As a result, $c/a$ is slightly underestimated by our theoretical prediction for laboratory vortices; the agreement is however excellent for numerical vortices.

\begin{figure}
	\centering
	\includegraphics[width=1\linewidth]{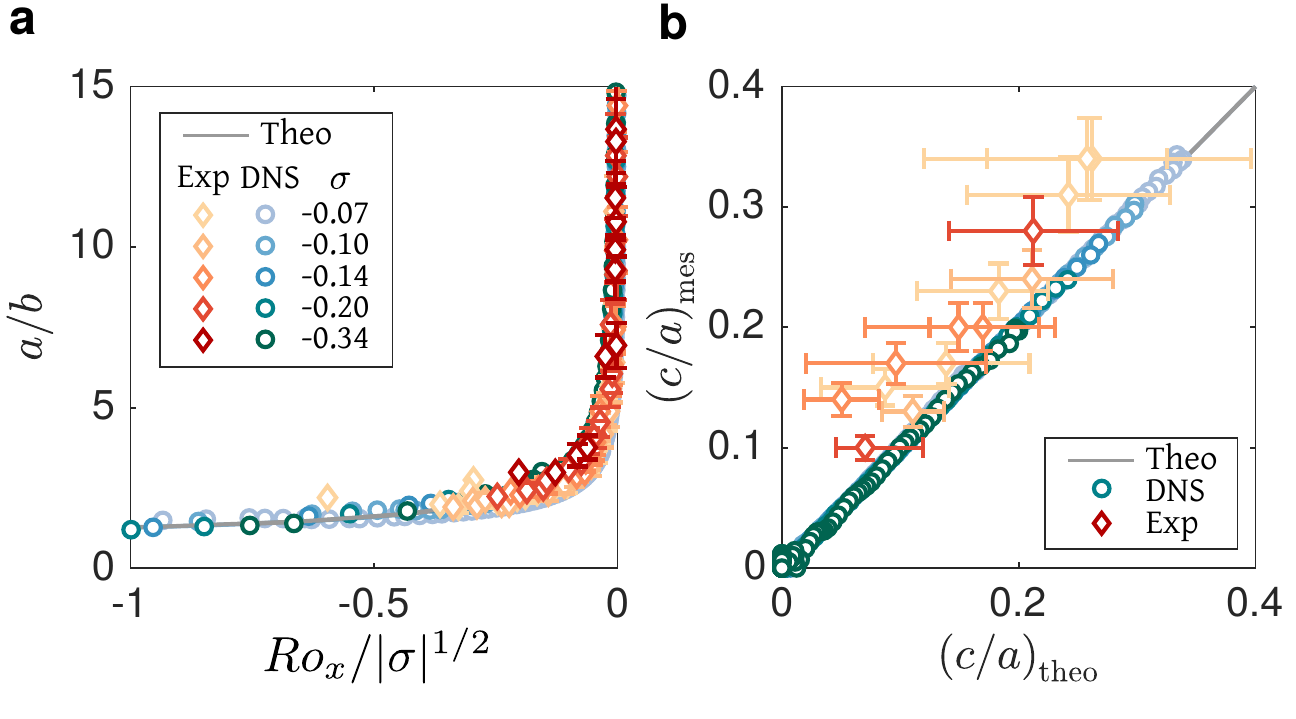}
	\caption{Evolution of the vortex shape for simulations (dots), experiments (diamonds) and theory (grey lines) for different background shear rates (colors). \textbf{a.} Horizontal aspect ratio $a/b$ of the vortices as a function of their Rossby number normalized by the shear rate ($Ro_x/\vert \sigma \vert^{1/2}$). Time increases from left to right since the Rossby number of a vortex decreases in absolute value by dissipation. Vertical error bars account for the variability when measuring the aspect ratios of the streamlines for a given velocity field (upper and lower bounds). Vertical and horizontal error bars are not represented when they are smaller than the markers. The theoretical prediction (equation \eqref{eq:vtx-horizaspect}) is plotted as a grey line, and involves no fitting parameter. \textbf{b.} Measured vertical aspect ratio $(c/a)_{\textup{mes}}$ as a function of its theoretical prediction $(c/a)_{\textup{theo}}$. Vertical error bars account for the variability when measuring the vertical aspect ratio of the vortex (upper and lower bounds). Horizontal error bars include uncertainties on the measured parameters ($Ro_x, \beta, N$ and $f$), but not on the vortex stratification $N_c$ which is not measured. We refer the reader to the Supplementary Information section 2 for details about the uncertainties. The theoretical prediction (equation \eqref{eq:vtx-vertaspect}) is plotted as a grey line, and involves no fitting parameter.}
	\label{fig:expnum-shape-evol}
\end{figure}

We now focus on Jovian vortices. In our model, the vortex shape results from a quasi-static equilibrium independent of the dissipation processes that govern the vortex decay: all that is requested is a time decoupling between the fast azimuthal motion controlling the equilibrium shape, and the slow dissipative processes controlling the long-term evolution. This time decoupling is valid for both our experimental and numerical vortices (see the dominant balances in Supplementary Information section 5.1), as well as for Jovian vortices (the GRS is at least 100 years old). 
We apply our laws to some of the most prominent Jovian anticyclones: the GRS in 1979 (as observed by the \textit{Voyager 1} mission), the Oval DE and BC in 1997 before their merger (\textit{Galileo}), and the Oval BA in 2007 (\textit{New Horizons}). Note that contrary to the GRS, the Oval BA was created recently after the merger of three White Ovals (FA, BC and DE) between 1998 and 2000. In 2007, it was thus only 7 years old, and yet this was long enough for it to evolve from the triangular shape that followed the  merger event to a classical elliptical shape\citep{choi_evolving_2010}. 

Our model requires four parameters: $Ro_x$, $\sigma$, $f$ and $N_c^2-N^2$. The Rossby number, the shear rate and the Coriolis parameter are known quite accurately. However, the picture is different for the stratification difference between the vortices and the atmosphere since the stratification inside any Jovian vortex has never been measured. Our estimation, based on thermal measurements, leads to lower bounds for our predicted vortex depths rather than absolute values. This point is discussed in the Methods section. The data, references, methods and uncertainties associated with each of these parameters are available in the Methods and Supplementary Information section 7.

\begin{figure*}
	\centering
	\includegraphics[width=\linewidth]{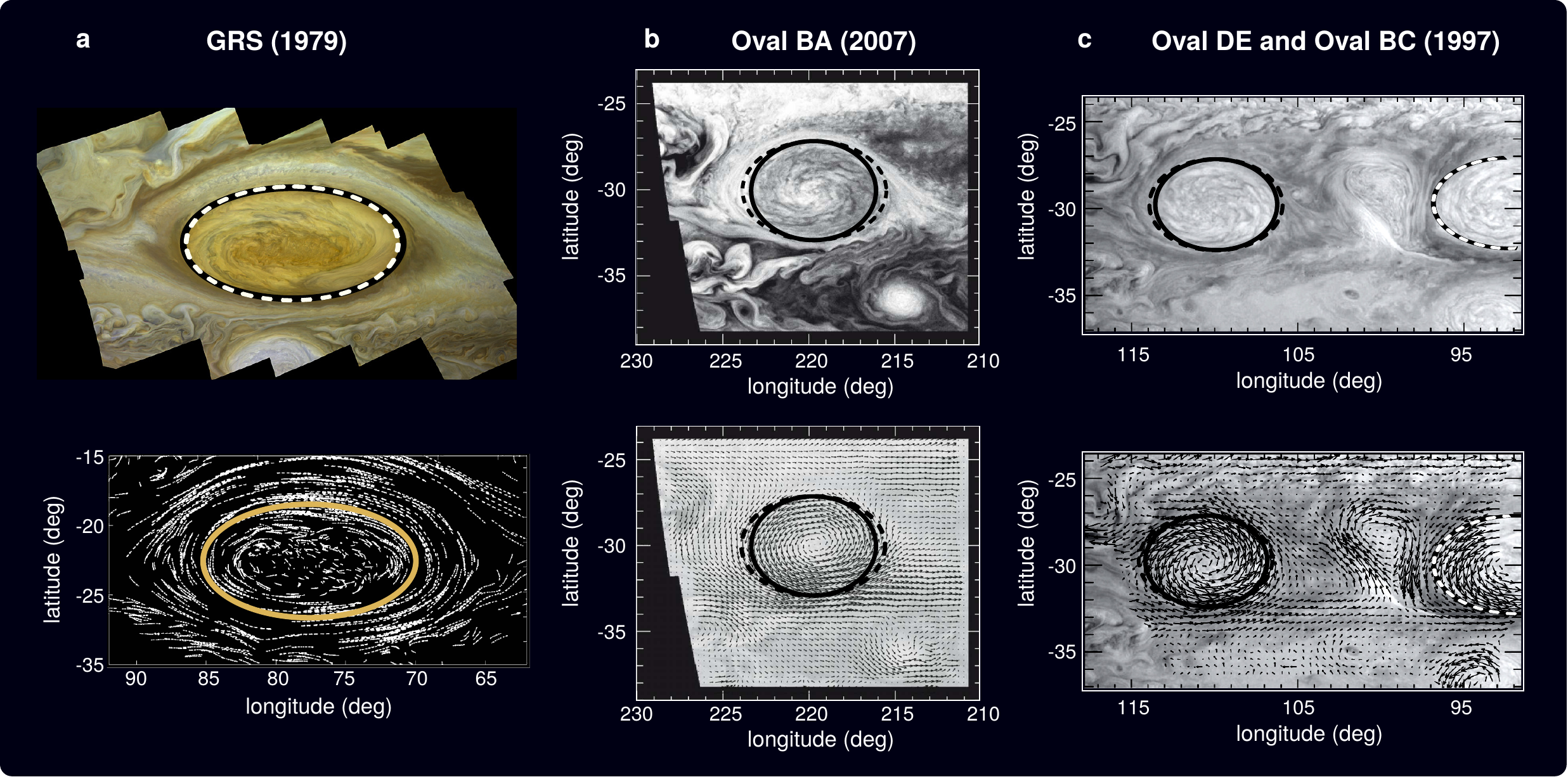}
	\caption{Images and velocity fields of Jovian anticyclones to which we superimposed ellipses with the measured aspect ratio (continuous lines) and the calculated one (dashed lines) using law (\ref{eq:vtx-horizaspect}) and the parameters reported in Table S2 of the Supplementary Information. \textbf{a.} Top: Mosaic of Great Red Spot (GRS) images (© NASA/JPL) taken by \textit{Voyager 2}\citep{johnson_bjorn_images_nodate} (reprinted with permission from Bjorn Jonsson). Bottom: Velocity vectors of the GRS as determined from \textit{Voyager 1} images\citep{shetty_interaction_2007} (© American Meteorological Society, used with permission). \textbf{b.} Oval BA as imaged by \textit{New Horizons} in February 2007 and associated wind vectors\citep{choi_evolving_2010} (reprinted with permission from Elsevier). \textbf{c.} Oval DE and BC as imaged by \textit{Galileo} in February 1997 and associated wind velocity vectors\citep{choi_evolving_2010} (reprinted with permission from Elsevier). Note that these two anticyclones merged between 1998 and 2000 with a third vortex (Oval FA) to give the current Oval BA\citep{choi_evolving_2010}.}
	\label{fig:vtx-ovalBA}
\end{figure*}

Applying relation (\ref{eq:vtx-horizaspect}), the predicted values for the horizontal aspect ratios for the GRS in 1979 (1.84 $\pm$ 0.14), the Oval BA in 2006 (1.45 $\pm$ 0.08) and the Oval DE and BC in 1997 (1.44 $\pm$ 0.14 and 1.67 $\pm$ 0.30) are of good order of magnitude and close to the measured ones at the cloud level (respectively 1.93, 1.22, 1.34 and 1.67). This  validates our approach and assumptions. These results are represented in Fig.\ref{fig:vtx-ovalBA} as ellipses superimposed to the vortices images and velocity fields. Contrary to their horizontal shape, the thicknesses of Jupiter's vortices are currently unknown. Some constraints are given by multi-layer quasi-geostrophic numerical simulations  \citep{vasavada_jovian_2005} which show that geostrophically balanced vortices tend to be baroclinically unstable if their thickness exceeds their width by a factor greater than $\sim f/N$. This leads to a maximum depth of $\sim 500$ km below the clouds for the GRS and the Oval BA. Later, it has been assessed that the large Jovians anticyclones should extend vertically down to the water cloud level \citep{de_pater_persistent_2010,wong_vertical_2011} (4-7 bar, i.e. 52 to 76 km below the clouds) which is consistent with the range of heights explored in numerical simulations \citep{legarreta_vertical_2008}. Our model predicts a half-height of $\sim$ $80^{+32}_{-16}$ km for the GRS. For the Oval BA, we find a half-thickness of $51^{+40}_{-13}$ km, $46^{+45}_{-12}$ km for the Oval DE and $64^{+56}_{-19}$ km for the Oval BC. These values are consistent with the estimated ones mentioned above and confirm the idea of shallow vortices which do not extend deeply into Jupiter's interior. 
In this view, the shallow vortices are embedded into deeper jets \cite{kaspi_jupiters_2018,cabanes_laboratory_2017} whose dynamics is independent of the anticyclones: once formed, the vortex decay is accompanied by a quasi-static equilibrium with the ambient shearing flow which governs their shape until they eventually disappear.

\begin{figure*}
	\centering
	\includegraphics[width=1\linewidth]{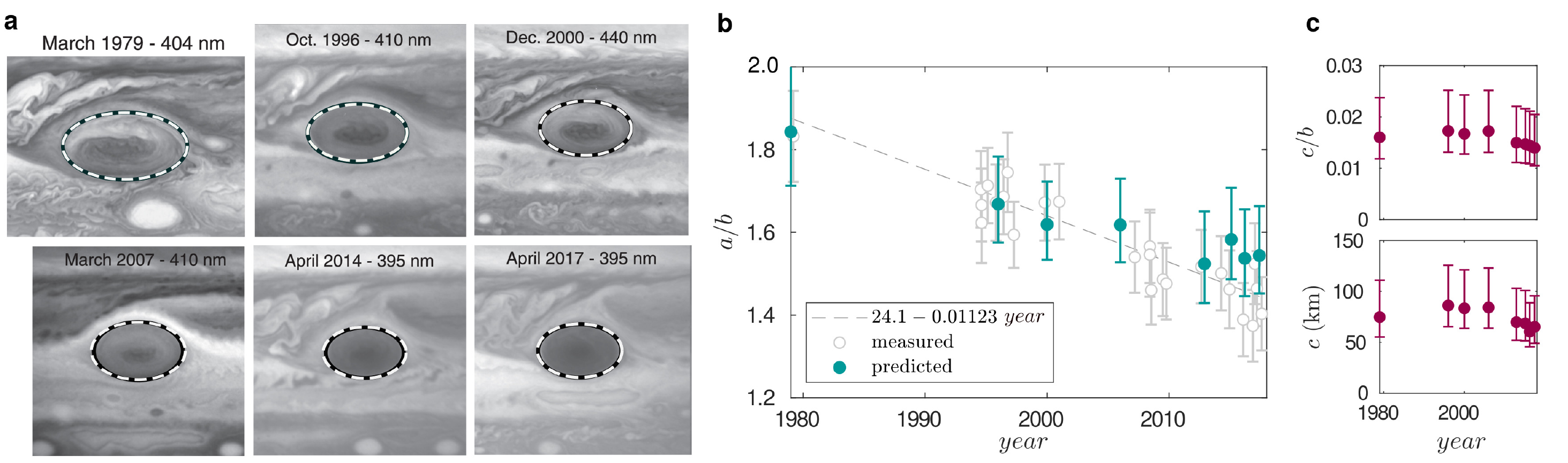}
	\caption{Observed and predicted evolution of the Great Red Spot (GRS) shape for the last 40 years. \textbf{a.} GRS appearance at blue/violet wavelengths\citep{simon_historical_2018} (© AAS, reproduced with permission), with the measured aspect ratio (black line) and the calculated one (white dashed line) using our model (equation \eqref{eq:vtx-horizaspect}) and the parameters reported in Table S3 of the Supplementary Information. \textbf{b.} Evolution of the GRS horizontal aspect ratio (major axis over minor axis on a horizontal plane) over the last 40 years. Measurements of the GRS aspect ratio are represented by the white dots\citep{simon_historical_2018}. Error bars correspond to uncertainties of 0.5$^\circ$ in latitude or longitude for each dimension\citep{simon_historical_2018}. The dashed line is the decreasing trend of the aspect ratio deduced from the measurements. The blue dots are the aspect ratio computed using our model (equation (\ref{eq:vtx-horizaspect})). Error bars account for the propagation of uncertainties of $\pm$ 10 m/s for the GRS velocities\citep{simon_historical_2018}  and $\pm$7 m/s for the zonal winds velocity \citep{shetty_interaction_2007}. \textbf{c.} Vertical aspect ratio $c/b$ and half-height $c$ as a function of time deduced from relation (\ref{eq:vtx-vertaspect}). To convert $c$ in kilometers, we use the measurements of $b$ given in Table 1 of Simon et al. (2018)\citep{simon_historical_2018}. Error bars account for uncertainties in velocities ($Ro_x$, $\sigma$), measured ellipticity ($\beta$) and stratification difference ($N^2-N_c^2$) which are given in Supplementary Information section 7.} 
	\label{fig:grs}
\end{figure*}

We now focus on the changes that occur in the GRS dynamics over the past 40 years. Spacecrafts data and \textit{Hubble Space Telescope} imagery show that the GRS is shrinking in the longitudinal direction (Fig.\ref{fig:grs}\textit{a}), decreasing from almost 35$^{\circ}$ extent in the late 1880s to less than 14$^{\circ}$ today \citep{simon_historical_2018}. The latitudinal extent of the GRS is also decreasing, but less rapidly, leading to a decrease in the horizontal aspect ratio\citep{simon_historical_2018} (dashed line in Fig.\ref{fig:grs}\textit{b}). The velocity field of the GRS has been measured at different times during this evolution showing an increase (in absolute value) in its longitudinal Rossby number. At the same time, the zonal wind velocities remained constant at the GRS latitude \citep{simon-miller_long-term_2010,tollefson_changes_2017}.  Assuming a constant shear rate, the predicted evolution of the horizontal aspect ratio according to our model agrees well with the measurements for the whole GRS evolution (blue dots in Fig. \ref{fig:grs}\textit{b}): for a given change in the longitudinal Rossby number, we predict the correct evolution of the horizontal shape, or conversely for a given shape evolution, we predict the correct evolution of the Rossby number. Note that if our quasi-equilibrium model is consistent with the recent evolution of the GRS, it does not give the physical mechanism responsible for this evolution. 

Finally, our model provides a remote access to the evolution of the GRS thickness for the past 40 years, which is not accessible with the available data. Using the calculated horizontal aspect ratio plotted in Fig.\ref{fig:grs}\textit{b}, the shear and the stratification difference reported in Table S2, and the Rossby numbers in Table S3, we compute the GRS vertical extent as represented in Fig. \ref{fig:grs}\textit{c}. Surprisingly, we find that the increase in absolute value of the longitudinal Rossby number compensates the decrease of the horizontal aspect ratio such that $c/b$ has remained constant through time. Since the latitudinal extent $b$ of the GRS has remained almost constant\cite{simon_historical_2018}, we conclude that the GRS has kept a mean half-thickness of $74^{+52}_{-28}$ km during its whole shrinkage.

In December 2017, preliminary results of the microwave radiometer (MWR) instrument onboard NASA's Juno spacecraft suggested that the GRS extends at least as deep as the instrument can observe, that is $\sim 300$ km below the cloud level\citep{greicius_nasas_2017}. However, this instrument measures thermal radiation, and the variations in brightness temperature can be interpreted as variations of opacity due to the abundance of chemical components such as ammonia, as well as variations in physical temperature \citep{janssen_mwr_2017}. Converting the MWR data into a signature of the density anomaly of the GRS is in our opinion a big interpretation step that requires further investigation. Since no scientific paper is for now published regarding these data, we leave this problem on standby. Nevertheless, if one assumes that the brightness temperature is entirely due to physical temperature variations, then what is measured is the extent of the density anomaly associated with the GRS. We argue that this density anomaly may have a vertical extent significantly bigger than the dynamical vertical extent of the vortex, that is the extent of the flow. We show in the Supplementary Information section 8 that if one uses the density anomaly to measure the vertical extent of the vortex, the latter could easily be $\sim$1.7 times what is measured using the winds. An observed density vertical extent of 300~km would thus give a dynamical vertical extent of 176~km consistent with our predicted range. Note that Juno flybys above the GRS allow gravity measurements among which the GRS signature will be detectable if the winds are deeper than $\sim$ 300 km \citep{galanti_determining_2019}. Upcoming measurements will thus challenge our model.

We conclude this study by pointing towards its limits and possible improvements. First, Jovian vortices exhibit a slight North-South asymmetry, barely visible in their shape, but apparent in their velocities \citep{choi_velocity_2007,choi_evolving_2010,shetty_changes_2010}. Including sources of asymmetry such as the $\beta$-effect and parametrizing deviation from ellipticity would improve the model's accuracy. Such effects could be tackled experimentally, with a sloping bottom to induce a topographic $\beta$-effect. However, we expect the influence on the vertical extent of those vortices to be negligible. Then, more evolved compressible models might lead to some changes of relevance for Jupiter's atmospheric dynamics. For instance, one could expect a vertical asymmetry of the density perturbation associated with the vortex. Additionally, small-scale time-dependent turbulence is present inside and outside Jovian vortices, but not in our laboratory model. The effects of such turbulence should also be tackled, even if it should generate only small perturbations of the potential vorticity anomaly associated with the vortex. Finally, as underlined by our long-term evolution discussion (Supplementary Information section 6), it would be interesting in the lab to set up a bulk shear rather than a boundary-driven one, which may lead to a more realistic interplay between the background and the vortex. We nevertheless argue that the results presented here, based on basic physics and first order balances, remain relevant and should be confirmed by up-coming Juno data.

\section*{Methods}
\footnotesize

\subsection*{Experimental set-up}
A plexiglass tank ($50 \times 50 \times 70$ cm) is filled with salt water linearly stratified in density using the double bucket method (see the resulting profile in Fig.\ref{fig:vtx-setup}\textit{c}). The tank is mounted on a table that rotates around a vertical axis at a rate $\Omega$. The buoyancy frequency is $N = 1 \pm 0.1 ~{\rm rad~s^{-1}}$ and the rotation rate is $\Omega = 0.5 \pm 0.05~{\rm rad~s^{-1}}$ such that $N/f = 1 \pm 0.2$. We impose a linear shear using a PVC belt encircling two co-rotating cylinders (Fig.\ref{fig:vtx-setup}). To create anticyclones in this gap, we inject through a capillary a volume of fluid having a constant density equal to the density at the injection height. Indeed, the geostrophic balance
\begin{linenomath*}
	\begin{equation}
	f \boldsymbol{e}_z \times \boldsymbol{u} = -\frac{1}{\rho} \nabla_h p,
	\label{eq:vtx-geostrophicbal}
	\end{equation}
\end{linenomath*}
where $\nabla_h p$ is the horizontal pressure gradient, implies that an over-pressure generates azimuthal velocities going in an opposite direction compared to the background rotation, i.e. an anticyclone ($Ro<0$). Additionally, relation (\ref{eq:vtx-vertaspect}) shows that $Ro \in [-1,0[$ (equilibrium anticyclonic motions) constrains $N_c < N$, where $N_c$ is the buoyancy frequency of the stratification at the core of the vortex. In other words, an anticyclone is under-stratified compared to the background density profile, that is why injecting a well mixed fluid is relevant. Note that the topographic $\beta$-effect resulting from the free-surface deformation due to rotation is negligible in our case. Velocity field measurements are performed in the equatorial plane of the vortex using particle image velocimetry (PIV). The computed velocity fields are used to measure the Rossby number and the equatorial ellipticity $\beta$ at each time during the slow vortex decay. To do so, we plot several streamlines near the vortex center and fit an ellipse to each of them. For some experiments we add a fluorescent dye in the injected fluid (Rhodamine B) to follow its evolution in a vertical plane. A detailed description of the experimental methods, parameters and uncertainties is available in the Supplementary Information (sections 2 and 4 and Table S1).

\subsection*{Direct numerical simulations (DNS)}
\label{sec:nummethod} 

We performed direct numerical simulations (DNS) to compare with our experimental results and to extend them to a wider range of parameters. To this aim, we solve the full system of equations (i.e. the continuity equation, Navier-Stokes equations in the Boussinesq approximation, and advection-diffusion equation of the stratifying agent) using the open-source spectral element solver Nek5000 \citep{fischer_nek5000_2008}. These equations are solved in a rectangular box to mimic the experimental setup. The boundary conditions are periodic in both the stream-wise ($x$) and vertical ($z$) directions. Rigid no-slip insulating boundaries are imposed in the cross-stream ($y$) direction to mimic the shear, i.e. velocity $\boldsymbol{u} = \mp \sigma y ~\boldsymbol{e_x}$ and no density anomaly gradient at $y = \pm 1$. Details about the numerical methods, the flow initialization and the complete list of the numerical parameters are available in the Supplementary Information (sections 3 and 4). Here, we focus on numerical simulations for which only the shear rate was changed and all the other parameters are fixed.

\subsection*{Parameters for Jovian vortices}

To apply our model to Jovian vortices, four parameters are required: the longitudinal Rossby number of the vortex $Ro_x$, the shear rate $\sigma$, the Coriolis frequency $f$ and the stratification difference between the vortex and the surrounding atmosphere $N_c^2-N^2$. The methods employed to estimate each parameter are provided in the next two subsections. The deduced parameters are reported in the Supplementary Information Tables S2 and S3. 

\subsubsection*{Velocities and length scales}

Horizontal length scales of Jovian vortices are measured based on wind velocities criteria for the GRS\cite{simon_historical_2018} and the Ovals BA and DE\citep{choi_evolving_2010}. For the Oval BC, we use a measurement based on cloud features\citep{mitchell_flow_1981}. From these data, we deduce for each vortex a measured horizontal aspect ratio and ellipticity to compare our predictions with (see Figs.4 and 5 and Supplementary Information Table S2).

To apply our model, the first quantity required is the longitudinal Rossby number $Ro_x$ of these vortices, that is the slope of the meridional velocity along an East-West profile, divided by the Coriolis frequency $f$. For the Oval BA and DE, we compute it by a linear fit on their meridional velocity profile at the core of each vortex, with and uncertainty of $\pm$ 5 m/s on the velocities\cite{choi_evolving_2010}. For the Oval BC for which we could not find velocity profiles, we use estimates of the North-South peak velocities\cite{mitchell_flow_1981} and divide them by the vortex semi-major axis length $a$. The resulting longitudinal Rossby numbers are given in the Supplementary Information Table S2.

For the GRS, we need to take into account the fact that it is a hollow vortex with a quiescent core. The detail of the velocity profile does not invalidate our approach since in the dynamical collar, we assume the same cyclo-geostrophic balance to hold, i.e. the pressure gradient compensates for the Coriolis and centrifugal forces arising from the non-zero azimuthal velocities. However, a correction needs to be added in the case of a hollow vortex to account for the fact that the distance from the core at which the velocity is maximal (the width of the vortex, $a$) is different from the characteristic distance of the pressure anomaly gradient (the width of the collar $a_c$)\cite{hassanzadeh_universal_2012}. The longitudinal Rossby number measured in the collar is $Ro_x= \frac{V_{max}}{a_cf} (1-\beta)$, where $V_{max}$ is the mean peak meridional velocity along an East-West profile. In that case, a prefactor $a_c/a$ should be added due to the centrifugal term for which it is the radius of curvature of the trajectory, i.e. the distance to the center that matters, not the size of the collar. Laws \eqref{eq:vtx-horizaspect} and \eqref{eq:vtx-vertaspect} are then modified as follow:

\begin{equation}
\beta^2 \left( 2 \frac{a_c}{a} \frac{Ro_x^2}{\sigma} +1 \right) + 2\beta  \left( \frac{a_c}{a} \frac{Ro_x^2}{\sigma} -1 \right) + 1 = 0,
\label{eq:vtx-horizaspect_hollow}
\end{equation}   
\begin{equation}
\left(\frac{c}{a}\right)^2 = \frac{Ro_x  \left[  1 + Ro_x~\frac{a_c}{a}~\frac{1+\beta}{1-\beta} \right] f^2 }{N_c^2 - N_f^2} 
\label{eq:vtx-vertaspect_hollow}
\end{equation}
where $Ro_x=\frac{V_{max}}{a_cf} (1-\beta)$ is the stream-wise Rossby number measured inside the collar. For the GRS in 1996, 2000 and 2006, we measure the longitudinal Rossby number by fitting meridional velocity profiles in the East-West direction inside its anticyclonic collar. The data are taken from Figure 5 in Choi et al. (2007) \cite{choi_velocity_2007} for 1996 and 2000, and Asay-Davis et al. (2009)\cite{asay-davis_jupiters_2009} for 2006, with an uncertainty of 10 m/s on the velocities \citep{simon_historical_2018} and 400 km of uncertainty for the measured distances $a$ and $a_c$. For the other dates, we use peak velocities and collar width measurements\cite{mitchell_flow_1981,simon_historical_2018}. The corresponding measured values for $a$, $a_c$ and $Ro_x$ are reported in the Supplementary Information Table S3. Regarding equation \eqref{eq:vtx-vertaspect_hollow}, rigorously speaking, the vertical aspect ratio is the aspect ratio between the pressure anomaly's vertical and horizontal characteristic length scales. Since to the best of our knowledge nothing is known about the influence of the GRS quiescent core on the density anomaly, we use the same assumption as for the other Jovian vortices, that is a pressure characteristic vertical scale equal to $c$. A complete and self-consistent model of the three-dimensional structure of a hollow vortex would be required, especially in terms of density anomaly, to conclude on the relevant scales. This lack of data and modeling leads us to use the simplest assumption, which is also the most consistent with our model, i.e. we assume that $a$ and $c$ are the semi-axes of the entire vortex.  To conclude on this point, note that although the quiet center of the GRS still remains today, it is significantly smaller than during the Voyager era (Supplementary Information Table S3). Additionally, no other vortices on Jupiter are known to have this hollow structure. They are rather very close to solid body rotation with a linear increase of the velocity in their core\citep{choi_evolving_2010} as assumed in our theoretical model, which hence seeks to be generic and applicable to the vast majority of Jovian anticyclones.

Additionally, our model requires estimates of the shear rate imposed by jets at the latitude of the vortices. Using linear fits on zonal winds profiles, we report those estimates and their errors for the GRS\citep{shetty_interaction_2007}, the Ovals DE and BC \cite{limaye_jupiter_1986} and the Oval BA \cite{tollefson_changes_2017} in the Supplementary Information Table S2.

\subsubsection*{Buoyancy and Coriolis frequencies}

The Coriolis parameter $f$, that is the amplitude of the vertical component of the rotation rate at the latitude of the vortices is taken from Table 3 of Mitchell et al. (1981) \cite{mitchell_flow_1981}.

The last but crucial parameter that we need to estimate is the difference of stratification between the vortex and the surrounding atmosphere $N^2-N_c^2$. To do so, we recall and discuss the method used in Aubert et al. (2012) \cite{aubert_universal_2012} supplementary material. The idea is to use temperature measurements that were performed in Jupiter's upper troposphere across the vortices and around them. Using the ideal gas equation and the fact that the pressure anomaly is zero at the top of the vortex ($z=h$), the density anomaly with respect to the ambient fluid at the top of the vortex can be expressed as $\Delta \rho /\rho = - \Delta T/T$, hence

\begin{equation}
\frac{T_a(z=h)-T_v(z=h)}{T_a(z=h)} = - \frac{\rho_a(z=h)-\rho_v(z=h)}{\rho_a(z=h)},
\end{equation}
where $T_a$ and $\rho_a$ are the temperature and density in the surrounding atmosphere, and $T_v,\rho_v$ within the vortex. At the core of the vortex ($z=0$), the density anomaly is zero, and a Taylor expansion leads to

\begin{equation}
\frac{T_a(z=h)-T_v(z=h)}{T_a(z=h)} \approx -h \frac{\left( \frac{\partial \rho_a}{\partial z} \right)_{z=0} - \left( \frac{\partial \rho_v}{\partial z} \right)_{z=0}}{\rho_a(z=h)} \approx \frac{h}{g} (N^2 - N_c^2).
\end{equation}
A crude estimation of the stratification difference between the vortex and the ambient can thus be obtained using temperature differences measurements: 

\begin{equation}
N^2-N_c^2 \approx \frac{g}{h} \left(\frac{T_a-T_v}{T_a}\right)_{z=h}.
\end{equation}

The temperature anomalies associated with the vortices have been measured quite accurately \citep{conrath_thermal_1981,flasar_thermal_1981,fletcher_thermal_2010}. Additionally, we adopt the pressure-temperature profile derived from the Galileo probe data \citep{seiff_thermal_1998} to obtain the mean atmosphere temperature at the measurement level. For the GRS, Figure 2 in Flasar et al. (1981) \citep{flasar_thermal_1981} shows a temperature anomaly of $8\pm1$ K at 50 mbars. With an atmospheric temperature at that level of $T_{a,50}=121\pm4$ K, we obtain a relative temperature anomaly of $(T_a-T_v)_{50}/T_{a,50} = 0.0661 \pm 0.0104$. For the Ovals DE and BC, Figure 1 in Conrath et al. (1981) \citep{conrath_thermal_1981} shows a temperature anomaly of $4\pm1$ K at $120\pm20$ mbars. With $T_{a,120}=115\pm2$ K, we obtain a relative temperature anomaly of $(T_a-T_v)_{120}/T_{a,120} = 0.0348 \pm 0.0093$. Since no thermal measurements were performed across the Oval BA, we make the assumption that its stratification is the same as the vortices from which it formed, hence we use the same value as for the Ovals DE and BC.

Finally, the distance $h$ between the measurement level and the vortex midplane where the temperature anomaly vanishes is also a poorly constrained parameter and should be considered with its uncertainties. The aforementioned anomalies are measured at 50 mbars ($z^*\sim+58$ km, $z^*=0$ being the 1 bar pressure level) for the GRS and 120 mbars ($z^*\sim+43$ km) for the Ovals. For the vortex midplane, the cold anomaly of the GRS was observed up to 500 mbar\citep{flasar_thermal_1981,fletcher_thermal_2010} ($z^*\sim+16$ km) meaning that the midplane (zero-anomaly) is located at higher pressures. According to observers, it could extend up to 2 bar  \citep{de_pater_persistent_2010}, that is $z^*\sim-20$ km. Consistently, in numerical modeling, the midplane of Jovian vortices is located between 400 to 1500 mbar\cite{morales-juberias_epic_2003,legarreta_vertical_2008}. If we take into account this large uncertainty, we obtain $h=60\pm18$ km for the GRS and $h=45\pm18$ km for the Ovals. With a gravitational acceleration of $g=23$ m s$^{-2}$ based on the Galileo probe measurements \citep{seiff_thermal_1998},  we finally obtain $N^2-N_c^2 = (2.53 \pm 1.16)\cdot 10^{-5}$ rad$^2$ s$^{-2}$ for the GRS and $N^2-N_c^2 = (1.78 \pm 1.19)\cdot 10^{-5}$ rad$^2$ s$^{-2}$ for the Ovals. The values are reported in the Supplementary Information Table S2 with all the parameters required to apply our model.  \\

Note that this method does not require an independent knowledge of the stratification in the atmosphere $N$ and within the vortex $N_c$, which is crucial since the stratification inside any of the Jovian vortices has never been measured. The drawback is that we use superficial measurements, and extrapolate them to deduce a density slope with the important assumption that this slope is constant. But contrary to $N_c$, the stratification of Jupiter's atmosphere has been measured and estimated (e.g. Galileo measurements\cite{seiff_thermal_1998} and modelling estimates extrapolating Voyager data\cite{legarreta_vertical_2008}). The result is that $N$ is not constant in the range of pressure considered here for the vortex midplane. In the upper troposphere, both Voyager data \cite[][Figure 2]{legarreta_vertical_2008} and estimates from inverse problems \cite{shetty_changes_2010} agree on $N \sim 0.02$ rad s$^{-1}$. At deeper levels in the atmosphere, this stratification is supposed to decrease and reach $N\sim0.005$ rad s$^{-1}$  for pressures between 1 to 7 bars \citep{legarreta_vertical_2008}. Unfortunately, we cannot rigorously take this decrease into account without knowing how the vortex stratification varies along with it since the essential parameter in our model is the difference between the stratification within the vortex and the ambient one, not the stratification itself. As such, one could ultimately reach the limit $N_c^2 \rightarrow  N^2$ for which the vortex vertical extent would become infinite. Our results thus depend on a proper estimate of the stratification difference with depth, and provide lower bounds for the vortex depths rather than absolute values.

With these parameters estimates, we can apply our model (laws \eqref{eq:vtx-horizaspect} and \eqref{eq:vtx-vertaspect}) to predict the ellipticity and the thickness of those Jovian anticyclones. The results are given in the main text and the Supplementary Information Table S2. 


\paragraph{Data availability} The data represented in Figs. \ref{fig:expnum-shape-evol}, \ref{fig:grs}b and \ref{fig:grs}c are available as Source Data 3 and 5. All other data that support the plots within this paper and other findings of this study are available from the corresponding author upon reasonable request.

\bibliography{Papier_Nature_2019}
\bibliographystyle{naturemag}

\paragraph{Supplementary information} accompanies this paper

\paragraph{Acknowledgments}
The authors acknowledge support from the European Research Council (ERC) under the European Union’s Horizon 2020 research and innovation program (grant agreement 681835-FLUDYCO-ERC-2015-CoG). This work was granted access to the HPC resources of Aix-Marseille Université financed by the project Equip@Meso (ANR-10-EQPX-29-01) of the program “Investissements d’Avenir” supervised by the \textit{Agence Nationale de la Recherche}.

\paragraph{Author contribution}
G.F. and M.LB designed research; D.L. and G.F. performed the experiments; B.F developed the numerical code; D.L. ran the numerical simulations; D.L., G.F., B.F. and M.LB analyzed the experimental and numerical data; D.L. led the writing of the paper.

\paragraph{Competing interest} The authors declare no conflict of interest.
\paragraph{Correspondence} Correspondence and requests for materials should be addressed to lemasquerier.pro@protonmail.com

\end{document}